\documentclass[singlecolumn,aps,pra,superscriptaddress]{revtex4-2}

\usepackage{graphicx}
\usepackage{amsmath}
\usepackage{amssymb}
\usepackage{dcolumn}
\usepackage{bm}
\usepackage{setspace}
\usepackage{color}
\usepackage[capitalize,nameinlink]{cleveref}
\usepackage{color,soul}

\begin{document}
\title{\LARGE Orientational dynamics of anisotropic colloidal particles in a planar extensional flow}
\author{{Dinesh Kumar}}
\email{dineshk2@illinois.edu}
\affiliation{ Department of Chemical and Biomolecular Engineering \\ University of Illinois at Urbana-Champaign, Urbana, IL, 61801}
\date{\today}

\begin{abstract}
Suspensions of anisotropic particles are commonly encountered in a wide spectrum of applications, including industrial and architectural coatings, targeted drug delivery and manufacturing of fiber-reinforced composites. A grand challenge in the field of chemical and material processing is robust production of strongly aligned fibers at the microscopic level, as this is routinely linked with enhanced mechanical properties at the macroscopic level. While the investigation of the microstructure of anisotropic colloids in shear flows has garnered a lot of theoretical and experimental attention, the case of extensional flow remains poorly understood due to several experimental challenges. In this article, we present a theoretical framework for predicting the steady and transient orientations of anisotropic particles in a flowing liquid undergoing precisely defined steady and time-dependent planar extensional flow at the stagnation point of a Stokes trap device. In particular, we analytically solve the Fokker-Planck equation for estimating the probability distribution function describing the orientation dynamics of rod-like objects as a function of flow strength (Peclet number, \textit{Pe}) and probing frequency (Deborah number, \textit{De}). The theoretical results are compared with recent experiments and reasonable agreement is found. We also discuss the challenges involved in obtaining a full closed-form solution for the transient dynamics of anisotropic particles in oscillatory time-dependent extensional flow. Overall, our theoretical framework provides a way to compare the orientation dynamics of rod-like particles with experiments that have been performed using a new experimental technique involving the Stokes trap and precise flow-control over the orientation of particles.\\
\end{abstract}

\maketitle

\section{Introduction}
There are numerous applications in science and engineering where it is useful to orient fibers, nanowires and nanotubes in a certain fashion to control the overall rheology and hydrodynamic characteristics of the suspension \cite{la_porta_optical_2004, mathai_simultaneous_2011,mathai_simultaneous_2013}. For instance, obtaining desired material properties of fiber-reinforced composites involves specific alignment of the particles to improve the overall mechanical, optical, electrical and magnetic properties \cite{leahy2017controlling, galajdaorientation_2003}. Such composites are routinely processed in complex flows through a combination of shearing and elongation. A first step towards understanding the processing of fiber composites is to study their dynamics in simple flows, such as shear and extensional flows, as they are relatively easy to model and can be replicated in laboratory experiments. The ability of extensional flows to strongly align the fibers along the stretching direction also makes them an attractive model flow-field as the composite is stiffer and stronger in this direction compared with any other direction. Hence, characterizing the flow-induced orientation dynamics of anisotropic particles through experiments and theory is of paramount importance in order to tailor the microstrucutre and desired properties with minimum cost and maximum reliability. Recently, the experimental results on transient orientation dynamics of a single rod-like particle was investigated \cite{kumar2019orientation} using the Stokes trap \cite{shenoy2019flow, kumar2020automation}.  Unfortunately, predicting the orientation dynamics of anisotropic particles in extensional flow has been challenging due to the complex interplay between the hydrodynamic forces exerted by the flow, and random thermal fluctuations in the suspending fluid causing Brownian rotational diffusion. Generally, the Fokker-Planck equation (FPE) accurately describes the time evolution of the particle orientation distribution function,$\psi(\phi,t)$, (or its moments) at an orientation, $\phi$, at time, \textit{t} :
\begin{equation}
\frac{\partial \psi\left ( \phi, t \right )}{\partial t}=D\frac{\partial^2 \psi\left ( \phi, t \right ) }{\partial \phi^2}-\frac{\partial \left [\omega\left ( \phi \right ) \psi\left ( \phi, t \right ) \right ]}{\partial \phi}
\end{equation}
where \textit{D} is the rotational diffusivity and $\omega$ is the rotational velocity of the anisotropic particle. The time evolution of $\psi(\phi,t)$ is of considerable interest to the fluid dynamics community as it determines the start up rheology of suspensions of anisotropic particles. However, it is well known that while the stationary solution at long times of the FPE can be given in closed form, full time-dependent solutions for an arbitrary $\omega(\phi)$ are extremely rare and the analytical solution exists for only a limited number of cases.\\

The orientation dynamics and distributions of rod-like particles has been extensively studied in shear flow through experiments and theory. Jeffery \cite{jeffery1922motion} investigated the suspension of non Brownian spheroidal particles in simple shear flow and found that there exists no steady state distribution as they periodically rotate in an unsteady motion known as Jeffery orbits. Hinch and Leal \cite{leal1972rheology} added the effect of rotational diffusion which led the distribution $\psi(\phi,t)$ to approach steady state and they made scaling arguments to qualitatively describe the evolution of $\psi(\phi,t)$ at high and low \textit{Pe}, where $Pe=\dot{\epsilon}/D$, and $\dot{\epsilon}$ is the imposed strain-rate by the shear flow. Recently, Leahy \textit{et al.} \cite{leahy2015effect} obtained an exact solution to the full Fokker-Planck equation (1) in orientation-space by an appropriate coordinate transformation to the phase-angle space under steady and time-dependent shear flows. An analytical solution of the orientation dynamics provided direct insights into the enhanced rotational diffusion due to shear. Leahy and co-workers published another paper \cite{leahy2017controlling} in 2017 where they used this analytical solution to describe the behavior of a dilute suspension of rod-like colloids under an arbitrary periodic, high-frequency shear flow. They presented optimized periodic waveforms to control the particle alignment and hence, the suspension rheology at an arbitrary moment in the flow-cycle. While the dynamics of rod-like particles has been extensively studied in shear-flows, it has not received much attention in the case of elongational flows. Another interest in the community is to compare the analytical approach of probability distribution functions with experiments under similar conditions. Since the demonstration of anisotropic colloids over long observation times in extensional flows has been challenging as the fluid elements separate exponentially in time \cite{kumar2020double,kumar2020conformational,kumar2016microfluidic}, the theoretical formulation of orientation dynamics in steady and time-dependent flows has not been attempted. \\

Motivated by recent experimental work which demonstrated arbitrary control over both the center of mass position and orientation of rod-like colloids using a Stokes trap \cite{kumar2019orientation,kumar2021shape,kumar2021nonlinear}, we aim to develop a theoretical framework for predicting the steady and transient orientations of anisotropic particles in a flowing liquid undergoing precisely defined steady and time-dependent planar extensional flow and to check whether these theoretical predictions agree with the experimental results.

\section{Quantitative Model and Governing Equations}
Let us consider a freely-suspended anisotropic particle placed at the center of a four-channel microfluidic cross-slot device having two alternating fluid inlets and outlets as shown in \textbf{Fig 1a} The slow, viscous and incompressible flow produced in the slot is well described in the Hele-Shaw regime by:
\begin{figure}
	\centering
	\includegraphics[width=\textwidth]{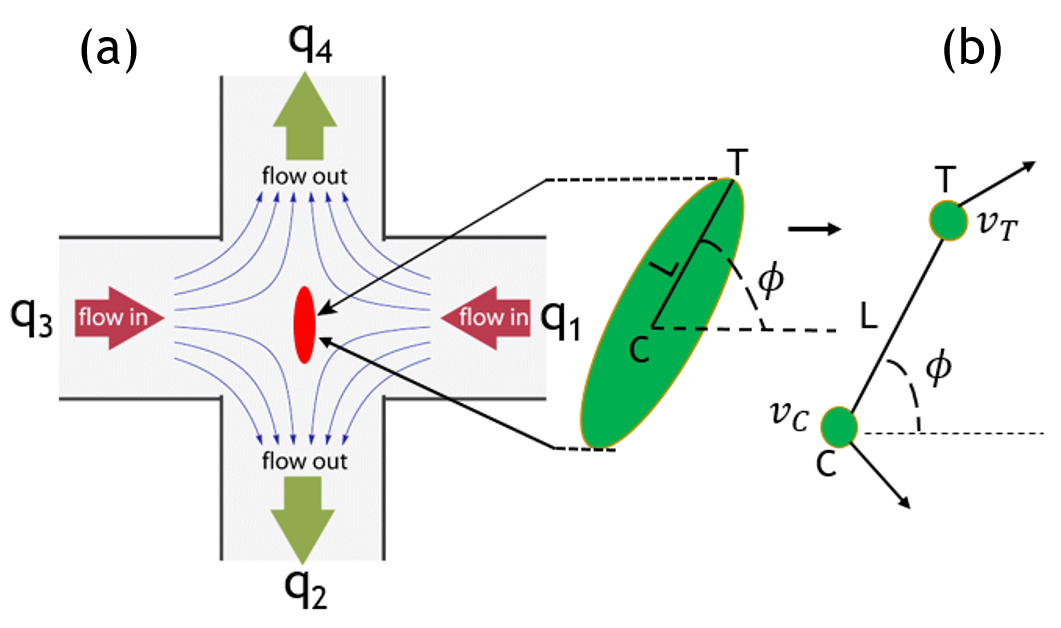}
	\caption{Schematic of the experimental set up. (a) Top view of the four channel microfluidic device. (b) The ellipsoid is modeled as two beads connected by a massless rigid rod.}
\end{figure}
\begin{equation}
\mathbf{ \dot{x}} = \frac{1}{\pi H}\sum_{i=1}^{4}\frac{(\mathbf{x}-\mathbf{R}_i)}{\left \| (\mathbf{x}-\mathbf{R}_i) \right \|^2}q_i
\end{equation}
where $\mathbf{x}$ is an arbitrary position, $q_i$ is a vector containing the flow rate through each channel and $\mathbf{R}_i=(R_{ix},R_{iy})$ is the position vector of $i^{th}$ channel and $H$ is the depth of the device. We model the anisotropic particle as two beads connected by a massless rod, where the lower bead is at the center of mass $\boldsymbol{x}_c=(x_c,y_c)$ of the rod, and the upper bead is at the extreme end, $\boldsymbol{x}_t$ of the major axis of the particle, being separated by a distance $L$. We can write the rotational velocity of the rod as: 
\begin{equation}
\mathbf{\dot{\phi}}=\omega(\phi)=\frac{1}{\pi H}\sum_{i=1}^{4}\frac{sin\left ( 2\phi \right )\|\mathbf{x}_c-\mathbf{r}_i\|^{2}-2cos\left ( 2\phi \right)\left ( x_c-R_{ix} \right )\left ( y_c-R_{iy} \right )}{\left \| (\mathbf{x}-\mathbf{r}_i) \right \|^4}q_i
\end{equation}
where $\phi$ is the angle made by the particle's major axis with the flow axis. We further assume that the center of mass position of the particle is kept fixed at origin of the device. This further simplifies equation (3) to: 
\begin{equation}
\dot{\phi}=-\dot{\epsilon} \sin\left (2\phi\right)
\end{equation}
where $\dot{\epsilon}=2q_1/\pi\textit{H}R_{cell}^2$ is the strain-rate imposed by the flow and $R_{cell}$ is the half-width of the channel.The corresponding probability distribution function (PDF) $\psi(\phi,t)$ of finding a rod of negligible cross-section at orientation $\phi$ and at time \textit{t} is given by the Fokker-Planck equation:
\begin{equation}
\frac{\partial \psi\left ( \phi, t \right )}{\partial t}=\frac{\partial^2 \psi\left ( \phi, t \right ) }{\partial \phi^2}+\textit{Pe}\frac{\partial \left [ \sin\left ( 2\phi \right ) \psi\left ( \phi, t \right ) \right ]}{\partial \phi}
\end{equation}

\begin{figure}
	\centering
	\includegraphics[width=\textwidth]{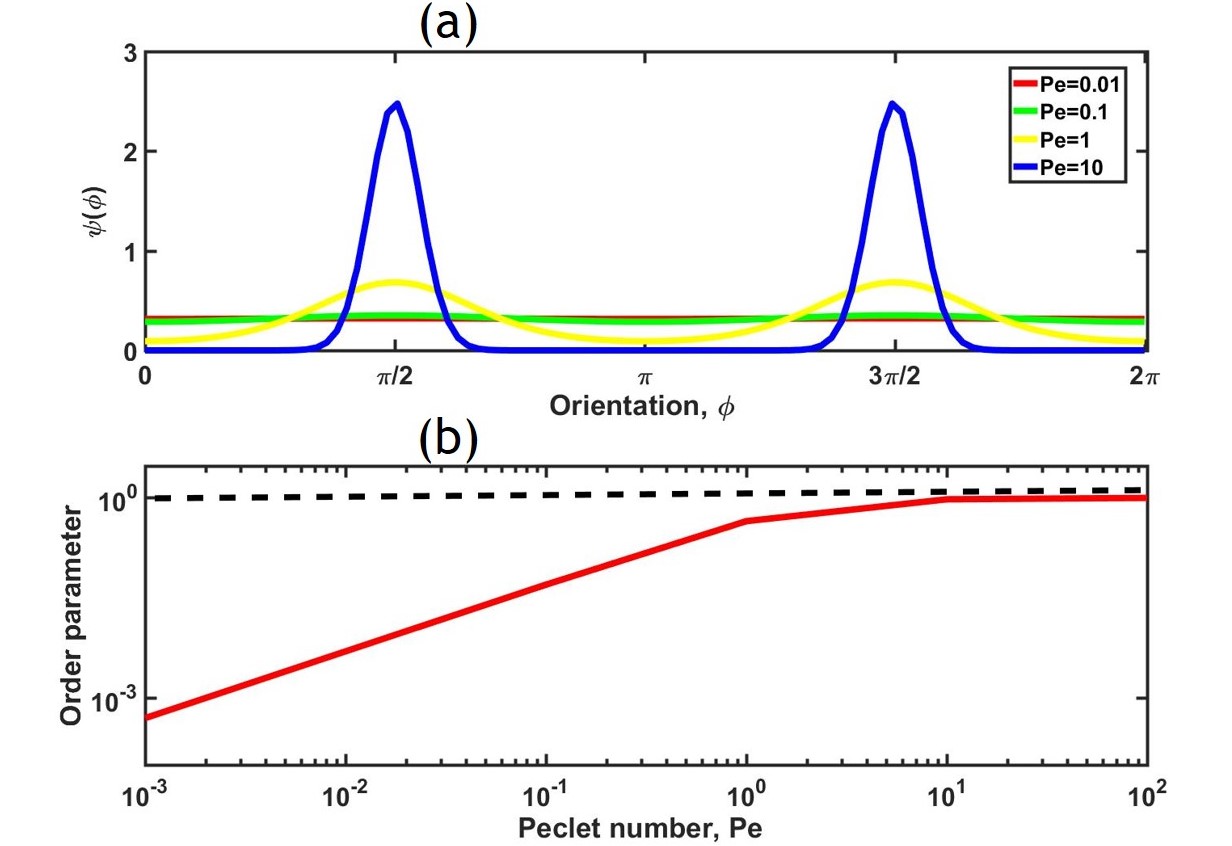}
	\caption{Stationary dynamics for the steady planar extension case. (a)The steady state PDF from equation (8) for Peclet number, Pe=0.01,0.1,1,10. (b). Value of Order parameter versus Peclet number as it approaches 1.}
\end{figure}

Since $\psi(\phi,t)$ is periodic with a period of $\pi$ as the rods are indistinguishable when oriented at $\phi$ or $\phi+\pi$, the boundary conditions and initial condition become:
\begin{equation}
\psi\left ( 0,t \right )=\psi\left ( \pi,t \right )=0 ;
\quad
{\psi}'\left ( 0,t \right )={\psi}'\left ( \pi,t \right ) =0
\end{equation}
\begin{equation}
\int_{0}^{\pi}\psi\left ( \phi,t \right )d\phi=1 ;
\quad
\psi\left ( \phi,0 \right )=\delta\left ( \phi-\phi_0 \right )
\end{equation}
where $\phi_0$ is the initial orientation angle of the particle.\\

\section{Results}

In this section, we present the analytical solution of equation (5) under certain cases:\\

\subsection{Steady state dynamics in continuous extension:}
At long times, the left hand side of equation (5), namely $\partial \psi\left ( \phi, t \right )$, equals zero and a closed form solution of the FPE is obtained as:
\begin{equation}
\psi\left ( \phi \right )=\frac{e^{-\textit{Pe}\cos2\phi}}{\int_{0}^{\pi}e^{-\textit{Pe}\cos2\phi}}
\end{equation}
The results we find for the long-time distribution function are displayed in \textbf{Fig. 2a} for small and large values of Peclet number ranging from 0.01 to 10. As the Peclet number \textit{Pe} increases, the orientation PDF becomes sharply peaked along the axis of extension $\phi=\pi/2$. Without going into the details, we also infer that the nature of differential equation in (5) changes for \textit{Pe} greater than 1000 as it becomes a singular perturbation problem in $\textit{Pe}^{-1}$. Hence, our analytical solution in (8) cannot predict the distribution function for $\textit{Pe}>1000$. To characterize the degree of alignment along the extension axis, we also define a 2D order parameter  $\textit{S}=2<\mathbf{p}\mathbf{p}>-\mathbf{\delta}$, where $\mathbf{p}$ is the unit vector along the major axis of the particle such that $\mathbf{p} = [cos(\phi),sin(\phi)]^{T} $. In the limit of $Pe\rightarrow\ 0$, the order parameter \textit{S} is zero and the distribution is termed isotropic. Also, at high Peclet number, the order parameter \textit{S} approaches unity and the distribution is strongly aligned along the extensional axis.\\

\begin{figure}
	\centering
	\includegraphics[width=\textwidth]{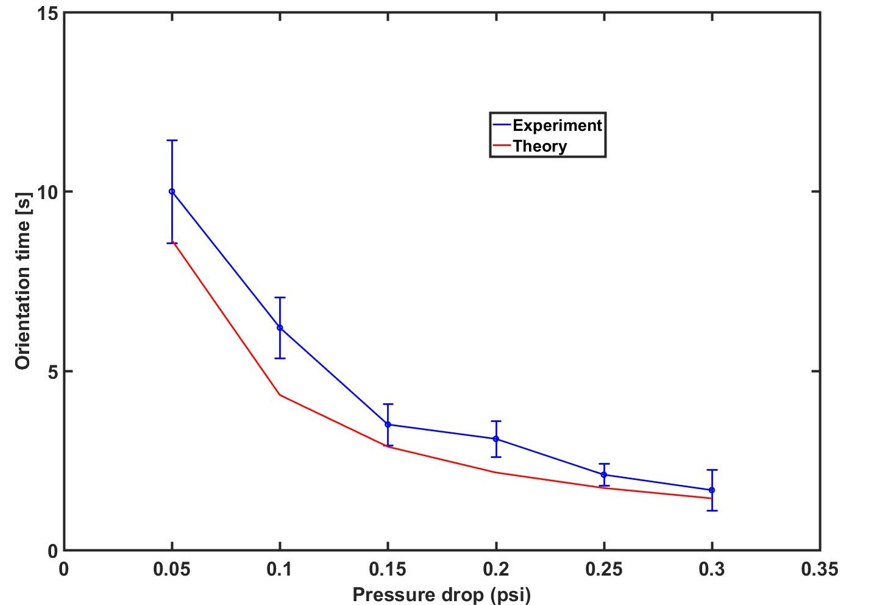}
	\caption{Characteristic orientation times for orientation angle changes from one angle to other as a function of pressure drop (strain-rate)}
\end{figure}

From Eq. (3), we can get the time evolution of $\phi(t)$ which is further used to predict the time taken by particle to change its orientation from an initial angle $\phi_0$ at time, $\textit{t}=0$ to a final orientation $\phi_f$ at time, \textit{t} as a function of strain-rate, $\dot{\epsilon}$. In the case where particle center of mass is not allowed to change, (3) reduces to (4) which is solved by integration through a simple separation of variable as the strain rate $\dot{\epsilon}$ is uniform at the center of cross-slot device. Finally, this yields:
\begin{equation}
t=-2\dot{\epsilon}\log\frac{\cot\phi_f}{\cot\phi_0}
\end{equation}

Also, the strain-rate $\dot{\epsilon}$ is linked with the pressure-drop $\Delta{P}$ across inlet and outlet of the cross-slot device through $\dot{\epsilon}=2q_1/\pi\textit{H}R_{cell}^2$ and $\Delta{P}=q_1/R$, where $R$ is the hydrodynamic resistance of the channel. \textbf{Fig. 4} depicts the time-prediction of change in the particle orientation from $\phi_0=89^{\circ}$ to $\phi_f=1^{\circ}$ as given in Eq. (9). To go further, we compared the recent experimental curve \cite{kumar2019orientation} to Eq.(9). We see a reasonable agreement between theory and experiments even though the particles experience perturbations when they reorient in the experiments which could change the strain-rate by a small magnitude. It is of note that the experimental curve describes the time-profile for an orientation change from $\phi_0=90^{\circ}$ to $\phi_f=0^{\circ}$ but slightly different values, $\phi_0=89^{\circ}$ and $\phi_0=1^{\circ}$ were chosen in theory because the trigonometric function $\cot\phi$ diverges near $\phi=0^{\circ}$.\\

\subsection{Transient dynamics in continuous extension :}
As mentioned earlier, a full closed-form solution to Eq. (5) does not exist. However, we can determine the transient dynamics by exactly solving (5) under the small angle approximation, $\sin2\phi \approx 2\phi$. This limits the choice of initial condition in Eq. (6) to $ \left |\delta\phi \right |\leqslant 15^{\circ}$ around the extensional axis $\phi=\pi/2$. Experimentally, this can be achieved by controlling the particle orientation to within $\pm 15^{\circ}$ along the extension axis at zero flow using the method described in \cite{kumar2019orientation} and gradually, increasing the applied flow. This can provide direct comparison with the transient dynamics obtained from theory, which we describe below.

\begin{figure}
\centering
\includegraphics[width=\textwidth]{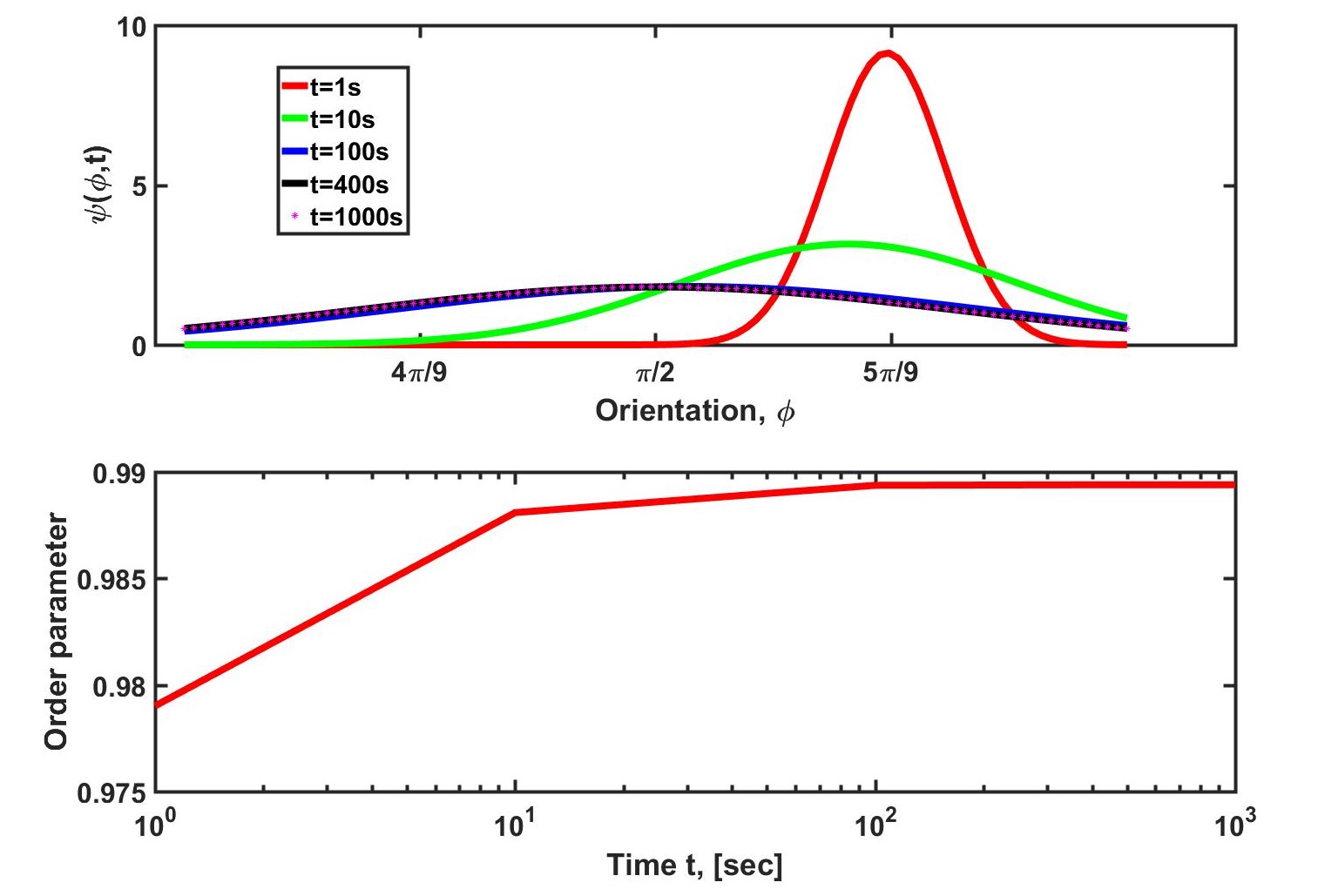}
\caption{Probability distribution function $\psi(\phi,t)$ versus $\phi$ at time t=1, 10, 100, 400 and 1000 s at $Pe=10.28$. Other parameters are $\phi_0=10^{\circ}$. The system reaches steady state at $t= 100$ s.}
\end{figure}

We solve Eq. (5) using a Fourier transform method and obtain the solution as: 
\begin{equation}
\psi\left ( \phi,t \right )=\sqrt\frac{Pe}{\pi\left ( 1-e^{-4\dot{\epsilon t}} \right )}\exp\left (-Pe \frac{\left ( \phi-\phi_0e^{-2\dot{\epsilon t}} \right )^2}{1-e^{-4\dot{\epsilon t}}} \right )
\end{equation}
At long times $t\rightarrow \infty$, Eq. (10) approaches  steady state and becomes similar to Eq. (8). For illustration, we plot the time-evolution of solution $\psi(\phi,t)$ and their corresponding order parameter \textit{S} as a function of time for two values of the Peclet number, $Pe=10.28 , 102.8$ in \textbf{Figs. 4 and 5}. The particle reaches steady-state corresponding to the onset of plateau in the plot of Order parameter versus time. As observed, the transient probability distribution function typically reaches steady state over a time scale of $\dot{\epsilon}^{-1}$.
\begin{figure}
	\centering
	\includegraphics[width=\textwidth]{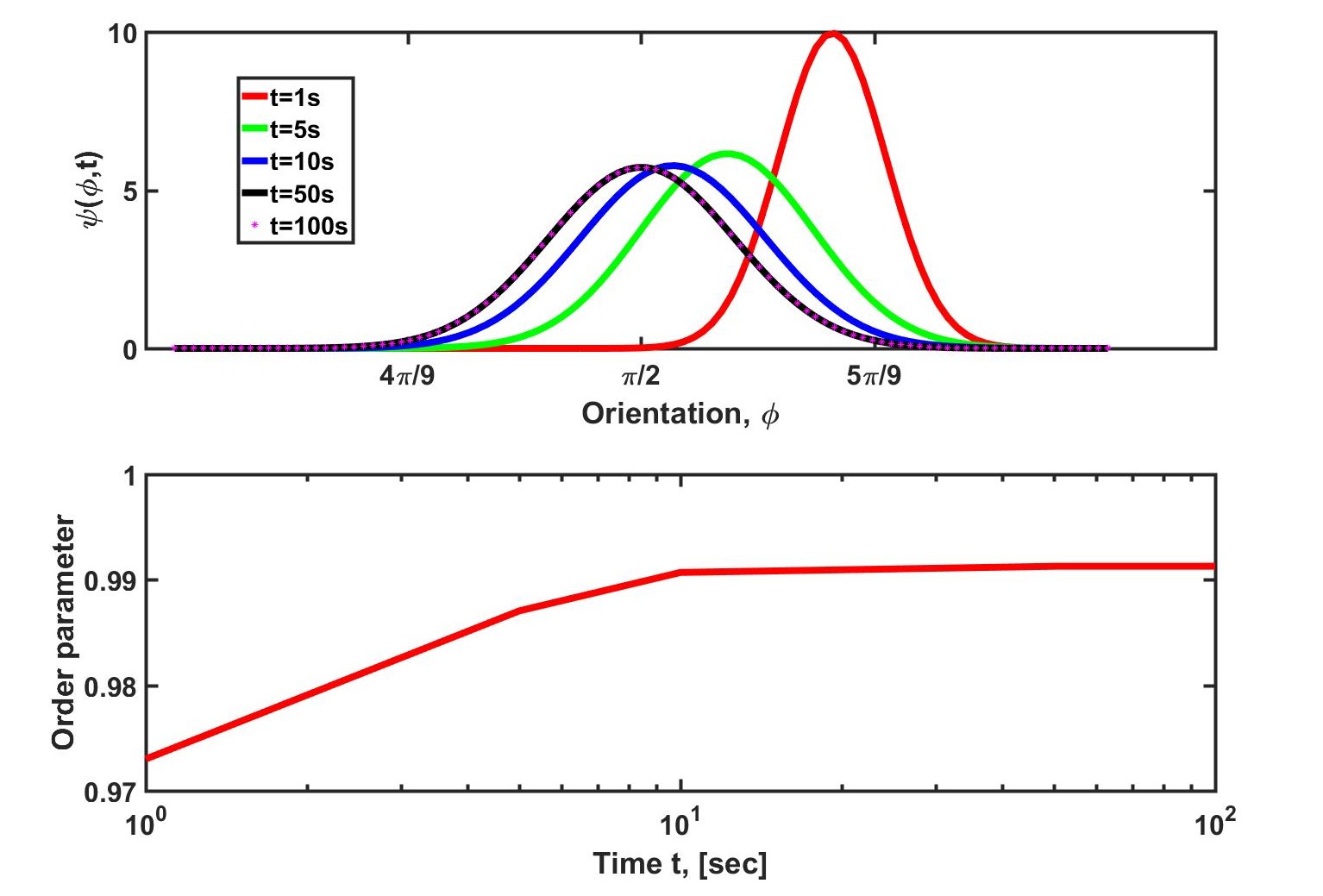}
	\caption{Probability distribution function $\psi(\phi,t)$ versus $\phi$ at time t=1, 5, 10, 50 and100 s at $Pe=102.28$. Other parameters are $\phi_0=10^{\circ}$.The system reaches steady state at a time of 50 s.}
\end{figure}\\

\subsection{Transient dynamics in oscillatory extension}
In this subsection, we aim to characterize the transient orientation distribution function $\psi(\phi,t)$ in large amplitude oscillatory extension (LAOE) over a wide range of Peclet number \textit{Pe} and Deborah number \textit{De} where $De=1/DT$ and \textit{T} is the cycle period of flow \cite{zhou2016transient}. As shown in \textbf{Fig.6}, we impose a sinusoidal strain-rate input $\dot{\epsilon}=\dot{\epsilon_0}\sin\left ( \frac{2\pi}{T} \right )$, where $\dot{\epsilon_0}$ is the maximum strain rate . The microdynamical equation:\\
\begin{equation}
tan\left (\phi_2 \right )=tan\left (\phi_1 \right )\exp\left ( -\frac{Pe}{\pi De}\left ( cos(2\pi DDet) \right ) -1\right )
\end{equation}
also allows us to plot the evolution of orientation $\phi$ as a function of sinusoidal strain in \textbf{Fig.6}. As expected, peak is observed for $\phi$ at $t=T/2$.
Next, we solve the FPE in Eq. (5) under the small amplitude approximation, and the result can be written as:
\begin{equation}
\psi\left ( \phi,t \right )=\frac{1}{\sqrt{2\pi A(t)}}\exp\left ( -\frac{(\phi-B(t))^2}{2A(t)} \right )
\end{equation}
where
\begin{equation}
B(t)=\phi_0\exp\left ( -\frac{-T\dot{\epsilon}}{\pi}\cos\left ( \frac{2\pi}{T} \right ) \right )
\end{equation}
\begin{equation}
A(t)=2D\int_{0}^{t}\exp\left ( -\frac{-T\dot{\epsilon}}{\pi}\cos\left ( \frac{2\pi}{T} \right ) \right )dt
\end{equation}
The solution of Eq. (11) is obtained through numerical integration by integrating over at-least five periodic cycles so that effect of initial flow transience have completely decayed out. The solution is then used to express the orientation dynamics in the context of two-dimensional \textit{(De,Pe)} space. It is noteworthy that our solution in Eq. (12) is only
\begin{figure}
	\centering
	\includegraphics[width=\textwidth]{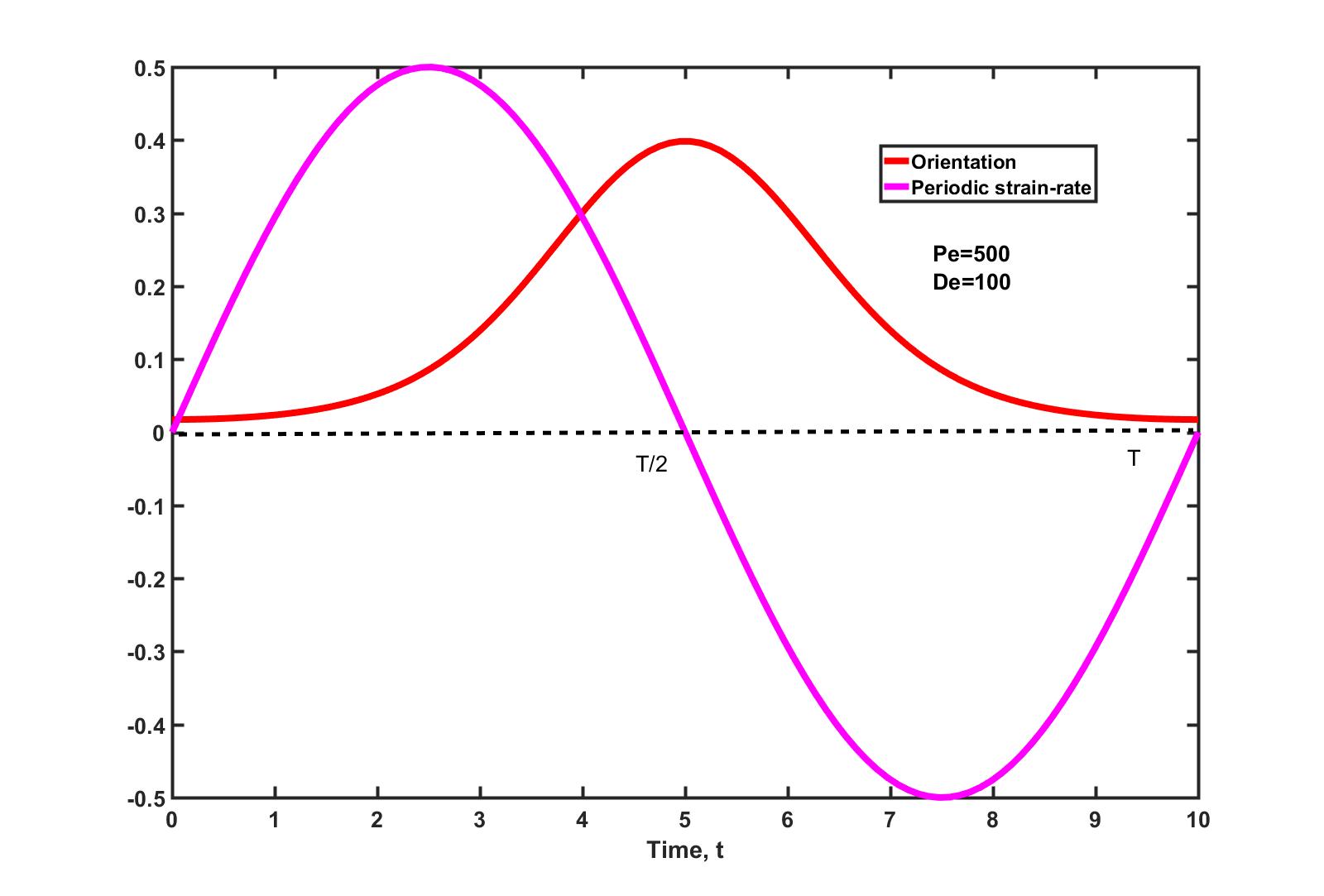}
	\caption{Schematic of sinusoidal strain-rate input for one full cycle (in magenta) and time evolution of orientation $\phi$ (in red) predicted from Eq.4}
\end{figure}
 valid for a set of $(Pe,De)$ at which the PDF $\psi(\phi,t)$ decays to zero beyond $ \pm{15^{\circ}}$ around the axis of extension $\phi=\pi/2$. Representative plots of order parameter in a full LAOE cycle are shown in \textbf{Fig.7} as a function of $Pe$ for oscillatory extensional flow $De>0$. It is observed that the order parameter in LAOE appears to be less on increasing $De$ (steady extension has higher order parameter compared to LAOE at the same imposed flow-strength) and reaches steady state at a higher $Pe$ for larger $De$. The complex interplay between $Pe$ and $De$ causes a decrease in order parameter due to presence of flow-oscillations at higher $De$.
 
\begin{figure}
	\centering
	\includegraphics[width=1\textwidth]{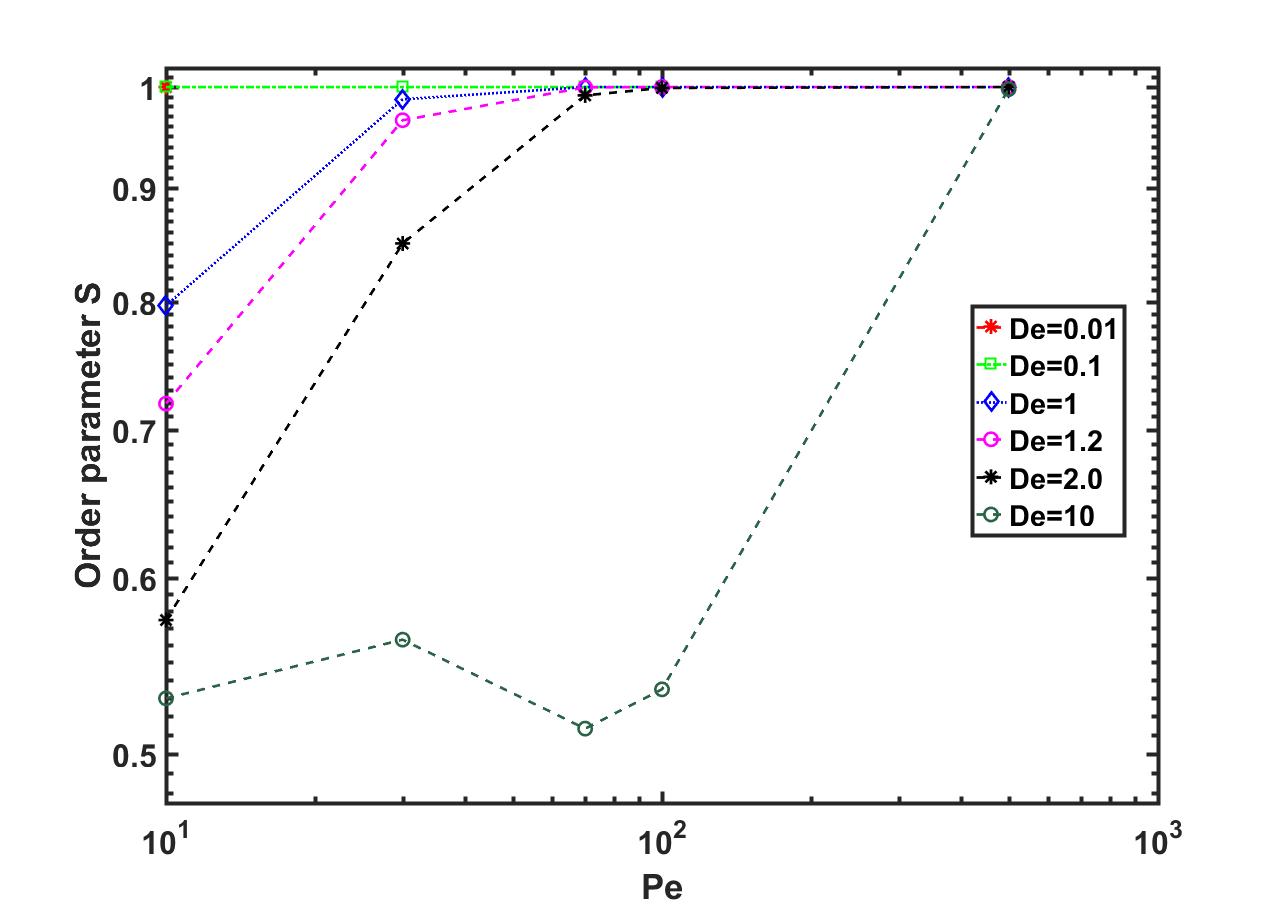}
	\caption{Order parameter as a function of Peclet number over a wide range of Deborah number}
\end{figure}

 Following the logic in \cite{zhou2016transient}, we next aim to construct a master curve for \textbf{Fig.7}. Our analytical solution is only valid in a small orientation space $[\pi/2-\delta\phi, \pi/2+\delta\phi]$ where $\delta\phi\approx15^{circ}$ and it cannot be used to define the average order parameter over the entire orientation space. However, it can be used to define a critical flow-strength at which transition from isotropic $S<1$ to aligned state $S=1$ happens since the transition occurs around a small angle around $\pi/2$. In plotting the master curve and hence, the dynamic oscillatory-flow based phase diagram, a key question is: What defines the critical Peclet number, \textit{Pe} at the boundary between  isotropic and aligned probability distribution function? While examining the steady-state distributions \textbf{Fig.2} at $\textit{De}=0$, we observed that order parameter \textit{S} saturates at a value of $S\approx0.996$. Using this observation, we define the transition from non-aligned to highly-aligned behavior to occur at a critical Peclet number $Pe_{crit}$ at which the order parameter $S=0.996$.

\textbf{Fig.8} shows a preliminary sketch of particle dynamic orientation distribution behavior in oscillatory extensional flow in the context of \textit{(De,Pe)} space. As discussed in \cite{zhou2016transient}, we classify four different regimes: (1) Regime I: Isotropic, quasi-steady state for $De<1$ and $Pe<Pe_{crit}$. Here, the period \textit{T} is large enough so that rods have sufficient time to relax through diffusion and achieve steady state. (2) Regime II: Isotropic, unsteady state for $De>1$ and $Pe<Pe_{cric}$ in which the rods do not get sufficient time in a cycle period to relax. (3) Regime III: Aligned unsteady state for $De>1$ and $Pe>Pe_{crit}$. (4) Regime IV: Aligned steady state for $De<1$ and $Pe>Pe_{crit}$. 
This phase diagram is derived from simulations and it can be experimentall verified by using the Stokes trap with large amplitude oscillatory extensional flow (LAOE) \cite{lin2021vesicle}. Since the analytical solution is based on an assumption that particles are found only in a small orientations-space around the axis of extension throughout the LAOE cycle, it will be interesting to look at the comparisons with the experiments. In the future, the validity of this assumption will be verified. In this respect, the results in this article provide an important first step in confirming that oscillatory extension reduces the alignment of anisotropic particles along extensional axis.

\begin{figure}
	\centering
	\includegraphics[width=\textwidth]{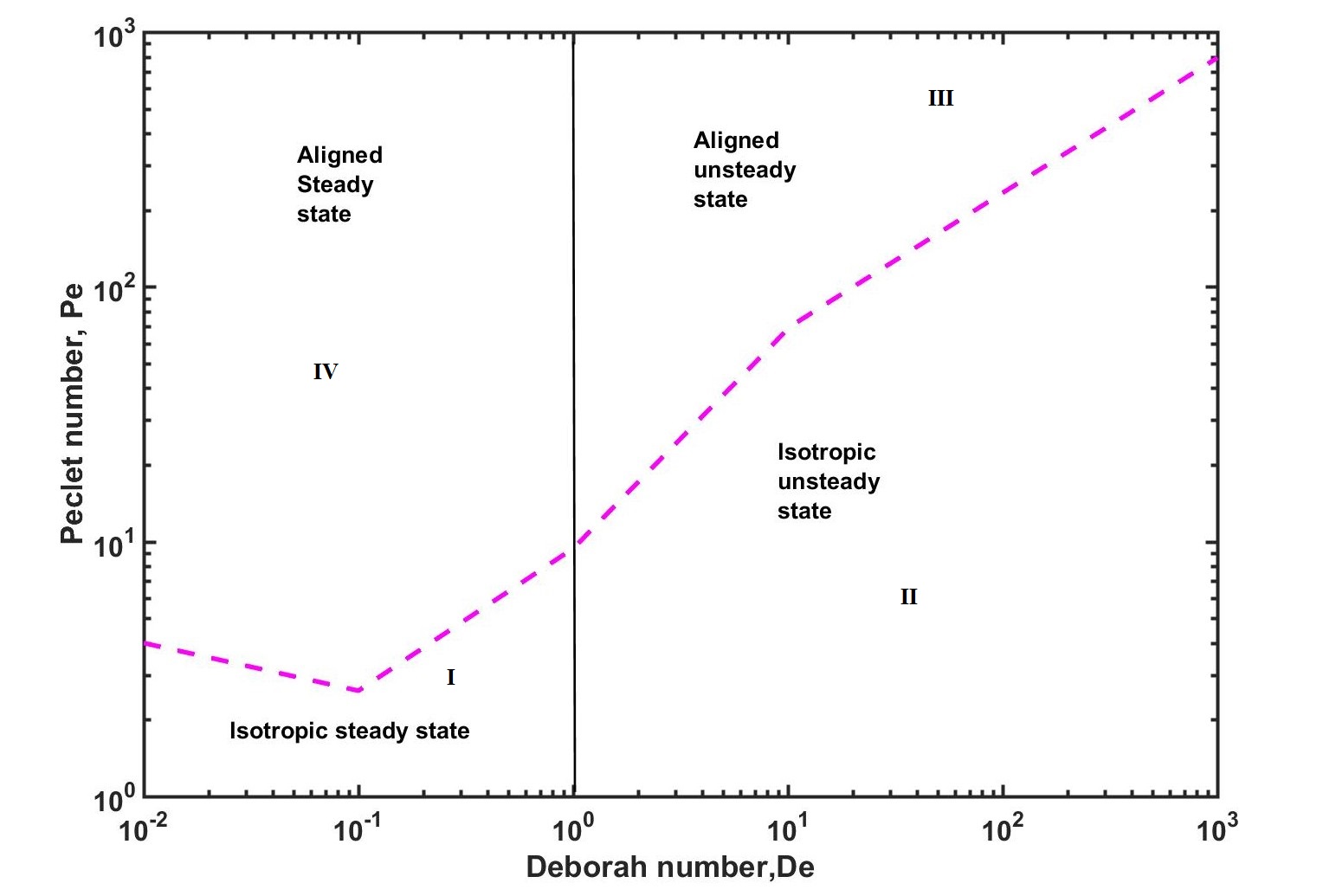}
	\caption{Average orientation dynamics of anisotropic particles in oscillatory extensional flow in the two-dimensional space of \textit{(De,Pe)}}
	\end{figure}

\subsection{Future experiments and simulations}
The results presented above provide a direct way to compare the stationary and dynamic orientation distribution function of rod-like particles with the recent experiments \cite{kumar2019orientation} that have been performed using a new experimental technique involving the Stokes trap and precise flow-control over the orientation of particles. The detailed predictions in this paper could be tested experimentally by precisely controlling the center of mass position of a single anisotropic particle using Stokes trap and directly observing the dynamics in steady flow. Also, our results on transient dynamics for a single rod-like particle in oscillatory extension can be matched by performing experiments in large amplitude oscillatory extension (LAOE) using Stokes trap, similar to \cite{lin2021vesicle,zhou2016transient}. 

In terms of modeling, the future work will involve completely solving the Fokker-Planck equation in Eq.5 by an appropriate series solution or perturbation method to establish the transient dynamics in complete orientation space. The effect of LAOE on the suspension stress and effective viscosity will be also be probed in the future through theory and experiments.

\section{Conclusion}
In this work, we investigated the stationary and transient dynamics of a single anisotropic particle in steady planar extensional and time-dependent oscillatory flow by analytically solving the full Fokker-Planck equation under the small angle approximation. Good agreement is found between the theoretical results and experiments for the case of dynamic evolution of the particle orientation as a function of strain-rate. Generally, steady-state distributions of particle orientations strongly peak along the extensional axis with increasing Pe. In oscillatory flow, the elongational axis rotates by an angle $\pi/2$ in each half-cycle and hence, the orientation of rod periodically changes to align with the stretching direction. We further characterized the average transient dynamics in LAOE in a two-dimensional \textit{(De,Pe)} space where the dynamic behavior was classified into four regimes based on values of \textit{Pe} and \textit{De} to predict the flow-based phase boundary for the transition from isotropic to an aligned state. Overall, our theoretical results establish new knowledge into our understanding of the non-equilibrium dynamics of single anisotropic particles in time-dependent flows and calls for  controlled experimental studies of rod-like particles in oscillatory extensional flows.

\bibliographystyle{unsrt}
\bibliography{rod_biblio}

\begin{thebibliography}{10}

\bibitem{la_porta_optical_2004}
Arthur La~Porta and Michelle~D. Wang.
\newblock Optical torque wrench: angular trapping, rotation, and torque
  detection of quartz microparticles.
\newblock {\em Phys. Rev. Lett.}, 92(19):190801, May 2004.

\bibitem{mathai_simultaneous_2011}
Pramod~P. Mathai, Andrew~J. Berglund, J.~Alexander Liddle, and Benjamin~A.
  Shapiro.
\newblock Simultaneous positioning and orientation of a single nano-object by
  flow control: theory and simulations.
\newblock {\em New J. Phys.}, 13(1):013027, 2011.

\bibitem{mathai_simultaneous_2013}
Pramod~P. Mathai, Peter~T. Carmichael, Benjamin~A. Shapiro, and J.~Alexander
  Liddle.
\newblock Simultaneous positioning and orientation of single nano-wires using
  flow control.
\newblock {\em RSC Adv.}, 3(8):2677--2682, January 2013.

\bibitem{leahy2017controlling}
Brian~D Leahy, Donald~L Koch, and Itai Cohen.
\newblock Controlling the alignment of rodlike colloidal particles with
  time-dependent shear flows.
\newblock {\em Journal of Rheology}, 61(5):979--996, 2017.

\bibitem{galajdaorientation_2003}
P{\'{e}}ter Galajda and Pal Ormos.
\newblock Orientation of flat particles in optical tweezers by linearly
  polarized light.
\newblock {\em Opt. Express}, 11(5):446--451, 2003.

\bibitem{kumar2019orientation}
Dinesh Kumar, Anish Shenoy, Songsong Li, and Charles~M Schroeder.
\newblock Orientation control and nonlinear trajectory tracking of colloidal
  particles using microfluidics.
\newblock {\em Physical Review Fluids}, 4(11):114203, 2019.

\bibitem{shenoy2019flow}
Anish Shenoy, Dinesh Kumar, Sascha Hilgenfeldt, and Charles~M Schroeder.
\newblock Flow topology during multiplexed particle manipulation using a stokes
  trap.
\newblock {\em Physical Review Applied}, 12(5):054010, 2019.

\bibitem{kumar2020automation}
Dinesh Kumar, Anish Shenoy, Jonathan Deutsch, and Charles~M Schroeder.
\newblock Automation and flow control for particle manipulation.
\newblock {\em Current Opinion in Chemical Engineering}, 29:1--8, 2020.

\bibitem{jeffery1922motion}
George~B Jeffery.
\newblock The motion of ellipsoidal particles immersed in a viscous fluid.
\newblock {\em Proc. R. Soc. Lond. A}, 102(715):161--179, 1922.

\bibitem{leal1972rheology}
LG~Leal and EJ~Hinch.
\newblock The rheology of a suspension of nearly spherical particles subject to
  brownian rotations.
\newblock {\em Journal of Fluid Mechanics}, 55(4):745--765, 1972.

\bibitem{leahy2015effect}
Brian~D Leahy, Donald~L Koch, and Itai Cohen.
\newblock The effect of shear flow on the rotational diffusion of a single
  axisymmetric particle.
\newblock {\em Journal of Fluid Mechanics}, 772:42--79, 2015.

\bibitem{kumar2020double}
Dinesh Kumar, Channing~M Richter, and Charles~M Schroeder.
\newblock Double-mode relaxation of highly deformed anisotropic vesicles.
\newblock {\em Physical Review E}, 102(1):010605, 2020.

\bibitem{kumar2020conformational}
Dinesh Kumar, Channing~M Richter, and Charles~M Schroeder.
\newblock Conformational dynamics and phase behavior of lipid vesicles in a
  precisely controlled extensional flow.
\newblock {\em Soft Matter}, 16(2):337--347, 2020.

\bibitem{kumar2016microfluidic}
Dinesh Kumar.
\newblock {\em A microfluidic device for producing controlled collisions
  between two soft particles}.
\newblock University of Toronto (Canada), 2016.

\bibitem{kumar2021shape}
Dinesh Kumar.
\newblock {\em Shape dynamics of anisotropic lipid vesicles using automated
  flow control}.
\newblock PhD thesis, 2021.

\bibitem{kumar2021nonlinear}
Dinesh Kumar and Charles~M Schroeder.
\newblock Nonlinear transient and steady state stretching of deflated vesicles
  in flow.
\newblock {\em Langmuir}, 37(48):13976--13984, 2021.

\bibitem{zhou2016transient}
Yuecheng Zhou and Charles~M Schroeder.
\newblock Transient and average unsteady dynamics of single polymers in
  large-amplitude oscillatory extension.
\newblock {\em Macromolecules}, 49(20):8018--8030, 2016.

\bibitem{lin2021vesicle}
Charlie Lin, Dinesh Kumar, Channing~M Richter, Shiyan Wang, Charles~M
  Schroeder, and Vivek Narsimhan.
\newblock Vesicle dynamics in large amplitude oscillatory extensional flow.
\newblock {\em Journal of Fluid Mechanics}, 929:A43, 2021.

\end{thebibliography}

\end{document}